\documentstyle[aps,psfig,preprint,tighten,floats]{revtex}

\begin{document}

\title{Effects of spinor distortion and density-dependent form factors
upon quasifree $^{16}$O$(\vec{e},e^\prime \vec{p})$}
\author{James J. Kelly}
\address{ Department of Physics, University of Maryland, 
          College Park, MD 20742 }
\date{\today}
\maketitle

\begin{abstract}
We propose an effective current operator for nucleon electromagnetic knockout
that incorporates spinor distortion and density-dependent nucleon form
factors using an effective momentum approximation.
This method can be used in a coordinate-space approach with either 
relativistic or nonrelativistic optical potentials and overlap functions.
We studied these effects for the $^{16}$O$(\vec{e},e^\prime \vec{p})$ 
reaction at $Q^2 = 0.8$ (GeV/$c$)$^2$.
Spinor distortion substantially enhances the left-right asymmetry while 
reducing the ratio between sideways and longitudinal recoil polarization 
for p-shell knockout by about $5\%$ for modest missing momenta.
We also find that the density dependence of nucleon form factors suggested 
by a quark-meson coupling model reduces the polarization ratio further.   
Much larger effects are obtained for the s-shell than for the p-shell.
However, both effects are subject to much larger Gordon ambiguities
than comparable nonrelativistic calculations.
\end{abstract}
\pacs{25.30.Dh,24.10.Jv,24.70.+s,27.20.+n}

\section{Introduction}
\label{sec:intro}

One of the central problems of nuclear physics is to determine the
sensitivity of hadronic properties to the local baryonic density.
For example, an early hypothesis, motivated in part by the EMC effect, 
was that the nucleon charge radius increases with density.
More recently, the quark-meson coupling (QMC) model has been used
to study the density-dependence of the nucleon electromagnetic form
factors \cite{Lu98a,Lu98b,Thomas98a} induced by coupling of their constituent 
quarks to the strong scalar and vector fields within nuclei.
However, because the predicted effects are relatively small at normal nuclear
densities, 
it will be very difficult to extract unambiguous results from measurements of 
cross sections for single-nucleon knockout from nuclei.
Fortunately, recoil polarization observables are expected to be much less 
vulnerable than cross sections to uncertainties in spectral functions, 
gauge ambiguities, and off-shell extrapolation of the single-nucleon current 
operator \cite{Kelly97}.
In the one-photon exchange approximation, the ratio between the longitudinal
and coplanar transverse polarization  transfers, $P^\prime_L / P^\prime_S$,
for quasifree kinematics is proportional to the ratio between electric and 
magnetic form factors, $G_E/G_M$ \cite{Arnold81}.
For large $Q^2$ and modest missing momentum,  
this relationship is relatively insensitive to final-state
interactions \cite{Kelly99a}.

Knockout calculations for quasifree kinematics are generally performed
using distorted-wave impulse approximations based upon either a relativized
Schr\"odinger or a Dirac equation.
Recent reviews of nucleon electromagnetic knockout can be found in 
Refs.\ \cite{Kelly96,Boffi93,Boffi96}.
Generally there are many differences between these approaches,
including choices of optical potential, overlap functions,  
current operator, and treatment of electron distortion.
We consider differences between optical potentials or overlap functions to
be superficial because it is always possible to transform a Dirac equation 
into an equivalent Schr\"odinger equation.
Furthermore, although many Schr\"odinger-based calculations 
\cite{Boffi80a,Giusti80} employ the
McVoy and van Hove \cite{McVoy}
nonrelativistic reduction of the electromagnetic current 
operator, we have shown that the effective momentum approximation permits
relativistic current operators to be used in relativized Schr\"odinger
calculations without need of nonrelativistic reduction 
\cite{Kelly96,Kelly99a}.  
Therefore, the  most important difference between these approaches is 
found in the dynamic enhancement of the lower components of Dirac spinors 
by scalar and vector potentials \cite{Hedayati-Poor95}.
We now extend the effective momentum approximation to include spinor
distortion and density-dependent nucleon form factors using a technique
based upon that of Hedayati-Poor {\it et al.}\ \cite{Hedayati-Poor95}
and investigate the effects of both bound-state and ejectile
spinor distortion on selected observables for proton knockout.
The operator which relates lower and upper components couples spin and
momentum, such that enhancement of the lower components modifies 
recoil polarization observables and enhances the left-right asymmetry for 
quasifree knockout.
However, we find that spinor distortion effects are quite sensitive to 
variations of the current operator arising from the Gordon identity.
Although these operators are equivalent on shell and differences for
nonrelativistic calculations are usually small, spinor distortion 
in relativistic calculations can produce substantial variations.

We have chosen to employ a coordinate-space representation because
that approach provides the most natural model for exploration of the
possible effects of medium modifications of nucleon form factors.
The effective momentum approximation is made to simplify the calculations
and is shown to be adequate for modest recoil momenta.
We find that substantial medium modification of recoil polarization
observables are expected, especially for s-shell knockout, using the
form factors from the QMC model.

\section{Model}
\label{sec:model}

\subsection{Spinor distortion}
\label{sec:spinor}

The derivation of the effective momentum approximation for nucleon
electromagnetic knockout reactions of the form 
$A(\vec{e},e^\prime \vec{N})B$
has been presented in considerable detail, 
including channel coupling in the final state, in Ref.\ \cite{Kelly99a}.
Thus, for the present purposes it suffices to employ a more schematic
notation in which state labels and channel coupling are omitted.
The nuclear matrix element has the basic form
\begin{equation}
 {\cal J}^{N\mu}({\bf q}) \approx 
\int d^3r_{NB} \;
  \exp{( i \bbox{\kappa} \cdot {\bf r}_{NA} )} \langle
  \chi^{(-)}({\bf r}_{NB}) | 
  \hat{J}_{\rm eff}^\mu( {\bf p}^\prime, 
                         {\bf p}^\prime - {\bf q} ) | 
  \phi({\bf r}_{NB}) \rangle  
\end{equation}
where ${\bf p}^\prime$ is the ejectile momentum, 
${\bf q}$ is the the momentum transfer,
${\bf p}_{m} = {\bf p}^\prime - {\bf q}$ is the missing momentum, 
${\bf r}_{NB}$ is the separation between the nucleon and the residual 
nucleus $B$, 
${\bf r}_{NA} = (m_B/m_A) {\bf r}_{NB}$ is the nucleon
position relative to the barycentric system,
and $\bbox{\kappa}$ is the momentum transfer in the barycentric frame.
The overlap between the $A$ and $A-1$ nuclei is represented by $\phi$, 
which is often called the {\it bound-state wave function},
while the ejectile distorted wave is $\chi$.
In the effective momentum approximation (EMA) we evaluate 
nucleon current operator
$\hat{J}_{\rm eff}^\mu( {\bf p}^\prime, {\bf p}^\prime - {\bf q} )$ 
using momenta determined by asymptotic kinematics in the laboratory frame, 
rather than operators, 
which then reduces the current to a matrix that acts upon nucleon spin.
Electron distortion is included in the EMA by adjusting the electron
momenta and flux  for acceleration by the Coulomb field, such that 
${\bf q} \longrightarrow {\bf q}_{\rm eff}$,
as described in Ref.\ \cite{Kelly99a}, but in the interests of notational
simplicity we omit the {\it eff} subscripts on ${\bf q}$, $\bbox{\kappa}$, 
and ${\bf p}_m$ because these corrections are small for the kinematics 
investigated here.
   
Suppose that a four-component Dirac spinor,
\begin{displaymath}
  \Psi( {\bf r} ) = \left( 
\begin{array}{c} \psi_+({\bf r}) \\ \psi_-({\bf r}) \end{array}
                    \right)
\end{displaymath}
where $\psi_+$ and $\psi_-$ are two-component Pauli spinors for positive
and negative energy components,
satisfies a Dirac equation of the form
\begin{equation}
\label{eq:Dirac}
 \left[ \bbox{\alpha} \cdot \bbox{p} + \beta(m+S) \right] \Psi
 = (E-V)\Psi
\end{equation}
with scalar and vector potentials $S$ and $V$.
It is well known that the upper component 
\begin{mathletters}
\label{eq:Darwin}
\begin{eqnarray}
 \psi_+ &=& D^{1/2} \phi \\
 D &=& 1 + \frac{S-V}{E+m}  
\end{eqnarray}
\end{mathletters}
can be obtained from 
an equivalent Schr\"{o}dinger equation of the form 
\begin{equation}
 \left[ \nabla^2 + k^2 - 2\mu \left( U^C + U^{LS} \bbox{L} \cdot 
\bbox{\sigma} \right) \right] \phi = 0 
\end{equation}
where
\begin{mathletters}
\begin{eqnarray}
  U^C &=& \frac{E}{\mu} \left[ V + \frac{m}{E} + \frac{S^2-V^2}{2E}
             \right] + U^D \\
  U^D &=& \frac{1}{2\mu} \left[ -\frac{1}{2r^2 D} \frac{d}{dr} (r^2 D^\prime)
      + \frac{3}{4} \left( \frac{D^\prime}{D} \right)^2 \right] \\
  U^{LS} &=& - \frac{1}{2\mu} \frac{D^\prime}{r D} \; .
\end{eqnarray}
\end{mathletters}
$D(r)$ is known as the Darwin nonlocality factor, and $U^C$ and $U^{LS}$ are
central and spin-orbit potentials; 
the Darwin potential $U^D$ is generally quite small.
Finally, because the lower component is related to the upper component by 
the spin-orbit coupling
\begin{equation}
\psi_- = \frac{ \bbox{\sigma}\cdot\bbox{p} }{E+m+S-V} \psi_+ \; ,
\end{equation}
we should anticipate that the strong Dirac scalar and vector potentials 
can affect the left-right asymmetry and recoil polarization observables 
for nucleon knockout.

Thus, we define an effective current operator as
\begin{equation}
\hat{J}^\mu_{\rm eff} = \tilde{\Lambda}_c({\bf p}^\prime,r) 
\gamma^0 \Gamma^\mu \Lambda_b({\bf p}_m,r) 
\end{equation}
where the subscripts $b$ and $c$ denote bound and continuum (ejectile) 
nucleons,
\begin{equation}
\Lambda_\alpha(\bbox{p},r) = \sqrt{ \frac{E_\alpha+m}{2m}} 
\left(  
\begin{array}{c} 1 \\
\frac{\bbox{\sigma}\cdot\bbox{p}}{E_\alpha+m+S_\alpha(r)-V_\alpha(r)}
\end{array} 
\right) D^{1/2}_\alpha(r) 
\end{equation}
represents a distorted spinor for $\alpha\in\{b,c\}$, and 
$\Gamma^\mu$ represents the electromagnetic vertex function.  
Therefore, the electronuclear matrix element becomes
\begin{equation}
\label{eq:Jeff}
 {\cal J}^{N\mu}({\bf q}) \approx 
\int d^3r_{NB} \;
  \exp{( i \bbox{\kappa} \cdot {\bf r}_{NA} )} \langle
  \chi^{(-)}({\bf r}_{NB}) | \tilde{\Lambda}_c({\bf p}^\prime,r_{NB}) \gamma^0
  \Gamma_{\rm eff}^\mu( {\bf p}^\prime, {\bf p}_{m} )
  \Lambda_b({\bf p}_{m},r_{NB})  | 
  \phi({\bf r}_{NB}) \rangle  
\end{equation}
where the vertex function and spinor distortion factors 
$\Lambda_\alpha(\bbox{p},r)$ are evaluated using the EMA.
The two-component bound-state and distorted waves are obtained as solutions 
to relativized Schr\"odinger equations using either intrinsically
nonrelativistic potentials or potentials derived from Dirac equations. 
Note that our previous results can be recovered simply by setting 
$S \rightarrow 0$ and $V \rightarrow 0$ in the lower components of 
$\Lambda_\alpha$ and by replacing $D^{1/2}\phi$ by a Woods-Saxon 
wave function.

Picklesimer {\it et al.} \cite{Picklesimer85,Picklesimer89}
first investigated effects of spinor distortion on response functions for 
proton knockout using a momentum-space formalism.
Several groups \cite{Boffi87,Jin94a,Udias95}
investigated the effect of Darwin nonlocality factors for the ejectile upon 
spectroscopic factors and missing momentum distributions using 
coordinate-space methods.  
Our present approach is based upon an effective current operator
proposed by Hedayati-Poor {\it et al.} \cite{Hedayati-Poor95}, 
who demonstrated the importance of potentials in the effective
current operator using an expansion in powers of $(E+m)^{-1}$.
Our previous calculations using the EMA are roughly equivalent
to carrying their ``nonrelativistic'' expansion to high order while
omitting potentials.
However, the present approach does not require such an expansion 
and is better suited to systematic investigation of the effects of spinor 
distortion and Gordon ambiguities in the vertex function.
Furthermore, we can also include channel coupling in final-state 
interactions using the formalism of Ref.\ \cite{Kelly99a} and
can include medium modifications of nucleon form factors by evaluating
the vertex function at the local density.

Given that $\frac{S-V}{2m} \sim -0.4$ for the bound state, one finds that 
the lower components are significantly enhanced in the interior.
In the present work we investigate  the effects of this dynamic enhancement 
of the lower components of distorted spinors upon observables for
nucleon electromagnetic knockout.
We refer to complete calculations based upon Eq.\ (\ref{eq:Jeff}) as SV, 
while calculations in which scalar and vector potentials appear in the 
lower components of just the bound state or just the ejectile spinor are 
identified as SVb or SVc, respectively.
Calculations using free spinors are labelled noSV.
In order to minimize superficial differences in momentum distributions or 
optical potentials,
we include the Dirac potentials in both Darwin factors, 
$D_b(r)$ and $D_c(r)$, whether or not those potentials are included in
the lower components of  $\Lambda_b$ or $\Lambda_c$.

For most calculations presented here, we employ the Dirac optical model
EDAD1 fitted by Hama {\it et al.} \cite{Hama90}
to proton-nucleus elastic scattering
for $A \ge 12$ and energies between 20 and 1040 MeV
and Dirac-Hartree wave functions obtained from the TIMORA code of
Horowitz {\it et al.} \cite{Horowitz86,TIMORA}.
These models also provide scalar and vector potentials for the bound state
and ejectile.
Alternatively, we can use nonrelativistic optical models and/or
binding potentials and obtain the necessary distortion factors by
exploiting the relationship between $S-V$ and the spin-orbit potential
$U^{LS}$, whereby 
\begin{equation}
\label{eq:D}
D(r)=\exp{\left(2\mu \int_r^\infty dr \; r U^{LS}(r) \right)} \; .
\end{equation}
This procedure has been used by Jin and Onley \cite{Jin94a} to compare
Darwin factors from a Dirac optical model with an optical potential based
upon our density-dependent empirical effective interaction (EEI) 
\cite{Kelly89b,Kelly94a}.

\subsection{Vertex functions}
\label{sec:Gordon}

The electromagnetic vertex function for a free nucleon can be 
represented by any of three operators \cite{deForest83,Chinn92}.
\begin{mathletters}
\label{eq:gamma}
\begin{eqnarray}
   \Gamma_1^\mu &=& \gamma^\mu G_M(Q^2) - 
     \frac{P^\mu}{2M}  F_2(Q^2)   
\label{eq:gamma1} \\
   \Gamma_2^\mu &=&  \gamma^\mu F_1(Q^2) + i \sigma^{\mu\nu}
     \frac{q_\nu}{2M}  F_2(Q^2)  \\
\label{eq:gamma2}
   \Gamma_3^\mu &=&  \frac{P^\mu}{2M}F_1(Q^2) + i \sigma^{\mu\nu}
     \frac{q_\nu}{2M} G_M(Q^2)  
\label{eq:gamma3}
\end{eqnarray}
\end{mathletters}
which are related by the Gordon identity.
Here $q=(\omega,{\bf q})$, $P=(E^\prime+E,{\bf p}^\prime+{\bf p})$, and
$Q^2 = {\bf q}^2 -  \omega^2$. 
Although $\Gamma_2$ is arguably the most fundamental because it is defined 
in terms of the Dirac and Pauli form factors $F_1$ and $F_2$,
one often uses $\Gamma_1$ because its matrix elements are simpler to
evaluate.
The third form is rarely used but is also equivalent on shell and, 
hence, is no less fundamental.

Unfortunately, because bound nucleons are not on shell, we require an
off-shell extrapolation for which no rigorous justification exists.
Lacking a better alternative, we employ the de Forest prescription
\cite{deForest83} in which the energies of both initial and final
nucleons are placed on shell, based upon effective momenta, while the
form factors are evaluated given $Q^2$ from the electron-scattering
kinematics.
Thus, we obtain the alternative prescriptions
\begin{mathletters}
\label{eq:gamma_bar}
\begin{eqnarray}
  \bar{\Gamma}_1^\mu &=& \gamma^\mu G_M(Q^2) - 
     \frac{\bar{P}^\mu}{2M}  F_2(Q^2)   
\label{eq:cc1} \\
   \bar{\Gamma}_2^\mu &=&  \gamma^\mu F_1(Q^2) + i \sigma^{\mu\nu}
     \frac{\bar{q}_\nu}{2M}  F_2(Q^2)  \\
\label{eq:cc2}
   \bar{\Gamma}_3^\mu &=& \frac{\bar{P}^\mu}{2M}F_1(Q^2) + i \sigma^{\mu\nu}
     \frac{\bar{q}_\nu}{2M} G_M(Q^2)  
\label{eq:cc3}
\end{eqnarray}
\end{mathletters}
where
\begin{eqnarray*}
   \bar{q} &=& (E^\prime - \bar{E}, {\bf q})   \\
   \bar{P} &=& (E^\prime + \bar{E}, 2{\bf p}^\prime - {\bf q})
\end{eqnarray*}
and where $\bar{E}=\sqrt{m^2 + ({\bf p}^\prime-{\bf q})^2}$ is placed on 
shell based upon the externally observable momenta ${\bf p}^\prime$ and 
${\bf q}$.

Finally, for the present calculations we employ MMD form factors \cite{MMD} 
and evaluate the current in the Coulomb gauge by modifying the longitudinal 
current to restore current conservation at the one--body level.
The sensitivity of nucleon knockout to alternative gauge prescriptions
was investigated in Ref.\ \cite{Kelly97}, but without spinor distortion.

\section{Results}
\label{sec:results}

Two experiments on the $^{16}$O$(e,e^\prime p)$ reaction were recently 
performed at Jefferson Laboratory using quasiperpendicular kinematics with
$\omega = 445$ MeV and $q=1000$ MeV, such that $Q^2 = 0.8$ (GeV/$c$)$^2$.
Experiment E89-003 \cite{TJNAF89-003} measured cross sections and separated 
the $R_T$ and $R_{LT}$ response functions for $p_m \lesssim 360$ MeV/$c$, 
while experiment E89-033 \cite{TJNAF89-033} measured helicity-dependent 
recoil polarization for $p_m = 85$ and 140 MeV/$c$.
The beam energy was 2.445 GeV and we include electron distortion in EMA.

The cross section for nucleon knockout can be expressed in the form
\begin{equation}
\label{eq:recoil-polarization}
 \frac{d^5 \sigma _{hs}}{d\varepsilon _{f}d\Omega _{e}d\Omega _{N}} =
   \sigma _{0} \left[1 + \bbox{P}\cdot \bbox{\sigma} 
   +h (A + \bbox{P}^{\prime} \cdot \bbox{\sigma})\right] 
\end{equation}
where $\varepsilon_i$ ($\varepsilon_f$) is the initial (final) electron
energy, $\sigma _{0}$ is the unpolarized cross section, $h$ is the
electron helicity, $s$ indicates the nucleon spin projection upon 
$\bbox{\sigma}$, $\bbox{P}$ is the induced polarization, $A$ is the electron
analyzing power, and $\bbox{P}^{\prime}$ is the polarization transfer 
coefficient.
Thus, the net polarization of the recoil nucleon $\bbox{\Pi}$ has two 
contributions of the form
\begin{equation}
   \bbox{\Pi} = \bbox{P} + h \bbox{P}^\prime
\end{equation}
where $\mid h \mid \leq 1$ is interpreted as the longitudinal beam 
polarization.
We choose to refer recoil polarization to a polarimeter basis in which
\begin{mathletters}
\begin{eqnarray} 
\label{eq:labpnuc}
\bbox{\hat{y}} &=& \frac{ \bbox{k}_i \otimes \bbox{k}_f }
                        {|\bbox{k}_i \otimes \bbox{k}_f|} \\
\bbox{\hat{x}} &=& \frac{ \bbox{\hat{y}} \otimes \bbox{p}_N }
                        {|\bbox{\hat{y}} \otimes \bbox{p}_N|} \\
\bbox{\hat{z}} &=& \bbox{\hat{x}} \otimes \bbox{\hat{y}} 
\end{eqnarray}
\end{mathletters}
where ${\bf k}_i$ (${\bf k}_f$) are the initial and final electron momenta
and ${\bf p}_N$ is the ejectile momentum in the laboratory. 
For coplanar quasiperpendicular kinematics with $\bbox{\hat{y}}$ upwards,  
it has become conventional to assign positive missing momentum to ejectile 
momenta on the large-angle side of ${\bf q}$, such that
$\theta_{pq}=\theta_p-\theta_q > 0$.

\subsection{Spinor Distortion}

The effect of spinor distortion upon differential cross sections is
shown in Fig.\ \ref{fig:sig_cc1}.
These calculations use the EDAD1 optical potential, Dirac-Hartree wave
functions, and the $\bar{\Gamma}_1$ vertex function.
Calculations which omit spinor distortion (noSV) are quite similar to
our previous results using Woods-Saxon bound-state wave functions and
the EEI optical model \cite{Kelly99a}.
The primary effect of distortion of the ejectile spinor is to increase
the differential cross section.  
The primary effect of distortion of the bound-state spinor is to
increase the left-right asymmetry
\begin{equation}
A_{LT}(\theta_{pq}) = 
\frac{ \sigma_0(-\theta_{pq}) - \sigma_0(\theta_{pq})}
     { \sigma_0(-\theta_{pq}) + \sigma_0(\theta_{pq})} \; .
\end{equation}
This effect is illustrated in Fig.\ \ref{fig:alt_cc1}, 
which shows that the enhancement of the left-right asymmetry increases 
with missing momentum.
This quantity provides particularly useful test of effective current
operators because for modest missing momenta it is relatively insensitive 
to details of the missing momentum distribution or to the choice of optical 
model.
Nevertheless, the structure at large angles is produced by optical model 
distortion and hence can be more variable.
The effects of spinor distortion are largest for the s-shell because the 
wave function is peaked where the Dirac potentials are strongest.

The sensitivity of spinor distortion to the choice of vertex function is 
illustrated in Fig.\ \ref{fig:sig_SV} for the cross section and in 
Fig.\ \ref{fig:alt_SV} for $A_{LT}$.
The largest effects are obtained with $\bar{\Gamma}_1$, especially for the
s-shell.
Gordon ambiguities in $A_{LT}$ are much larger when spin distortion 
factors are applied than those shown in Ref. \cite{Kelly97} without them. 

To assess the model dependence of the relationship between recoil polarization
and nucleon form factor, it is useful to define a polarization ratio as
\begin{equation}
r_{xz} = P^\prime_x / P^\prime_z
\end{equation}
which for a free nucleon at rest is proportional to $G_E/G_M$.
We can then compare $r_{xz}$ for a particular model either to a 
plane-wave calculation or to a baseline optical-model calculation.
In Fig.\ \ref{fig:rxz_SV} 
we chose $\bar{\Gamma}_2$ without spinor distortion as the
baseline calculation and display ratios between $r_{xz}$ with spinor
distortion to that baseline calculation.
At the kinematics of E89-033 we find that the effect of spinor distortion 
is approximately a 5\% reduction of $r_{xz}$ for p-shell knockout, with a 
$\pm 5\%$ Gordon ambiguity, while the s-shell effect is approximately $10\%$.  
The Gordon ambiguity increases rapidly with missing momentum, especially
for the s-shell, and is much larger with spinor distortion than without.

The roles of bound-state versus ejectile spinor distortion are compared
in Fig.\ \ref{fig:rxz_cc2}.
It is interesting to note that whereas the bound-state effect was dominant
for $A_{LT}$, the ejectile effect tends to dominate for $r_{xz}$.
Therefore, consistency between these observables, 
which sample spinor distortion rather differently,
provides a stringent test of the effective current operator.
Given that Gordon ambiguities are comparable in magnitude to the effects
predicted by models of the density dependence of nucleon form factors,
one must require consistency before attempting to infer a medium modification
of $G_E/G_M$. 

\subsection{Density dependence}

The possible effects of density dependence of nucleon form factors
upon the polarization ratio are illustrated in Fig.\ \ref{fig:rxz_QMC}.
The density dependence of the form factors was estimated by scaling
each Sachs form factor from the MMD model by the corresponding ratio 
between QMC form factors at density $\rho$ and in free space 
according to
\begin{equation}
G(Q^2,\rho) = G(Q^2) \frac{G_{\rm QMC}(Q^2,\rho)}{G_{\rm QMC}(Q^2)}
\end{equation}
The QMC form factors were computed using a bag radius of 0.8 fm
by Lu {\it et al.}\ \cite{Lu98a}.
These density-dependent form factors were evaluated at the local
density $\rho(r_{NB})$ and used in the coordinate-space current
matrix element given by Eq. (\ref{eq:Jeff}).
We find that at the kinematics of E89-003, $r_{xz}$ for the p-shell
is reduced by an additional factor of about 0.92 when density dependent
form factors are included, so that the net reduction relative to a
standard nonrelativistic DWIA calculation with free form factors is
about 0.88 for $p_m \approx 85$ MeV/$c$.
For the s-shell the net suppression of $r_{xz}$ is about 0.6 at
$p_m \approx 85$ MeV/$c$ and 0.25 at 140 MeV/$c$, with both the
relativistic and the density-dependent effects being much stronger
because higher densities dominate.
The effects on cross sections and $A_{LT}$ are much smaller and omitted.
These results suggest that an upcoming experiment \cite{TJNAF93-049} on 
$^4$He$(\vec{e}, e^\prime \vec{p})$ should be able to distinguish
density dependence from relativistic effects by measuring both $A_{LT}$
and $r_{xz}$ symmetrically about quasifree kinematics.

\subsection{Comparison with momentum-space approach}
\label{sec:momentum-space}

It is useful to compare our coordinate-space approach with the
momentum-space approach developed by Picklesimer, van Orden, and Wallace
\cite{Picklesimer85,Picklesimer87,Picklesimer89}
in which the current matrix element is expressed in the form
\begin{equation}
{\cal J}^\mu({\bf q}) = \int \frac{d^3 p}{(2\pi)^3} \;
\bar{\psi}^{(-)}({\bf p}^\prime,{\bf p}) \Gamma^\mu({\bf p},{\bf p}-{\bf q}) 
\phi({\bf p} - {\bf q})
\end{equation}
where ${\bf q}$ is the laboratory momentum transfer and
$\phi$ and $\psi$ are initial and final Dirac momentum-space wave functions 
for the struck nucleon. 
This approach avoids the effective momentum approximation, but does not
account for target recoil or include the off-shell extrapolation of the 
vertex function proposed by de Forest and customarily used to analyze data.
Nor does this approach provide for density-dependent form factors.

The original calculations by Picklesimer and van Orden used an optical
potential constructed by folding the nonrelativistic Love-Franey (LF)
effective interaction \cite{LF81}
with a simple three-parameter Fermi density distribution 
for $^{16}$O and using the prescription of Hynes {\it et al.}\ \cite{Hynes85}
to import this potential into the momentum representation of the Dirac 
equation.
We performed a similar calculation by using the same nonrelativistic optical
potential in the Schr\"odinger equation and Eq. ({\ref{eq:D}) to
generate the spinor distortion factors.
We also used the same bound-state wave functions, vertex function, and 
form factors and omitted electron distortion in order to emulate the
van Orden calculation as closely as possible in our approach.
Probably the most important remaining differences that might affect 
this comparison are
the effective momentum approximation in our coordinate-space approach
and the absence of recoil in the momentum-space approach,
but the relative importance of these aspects of the calculations is
not yet known.

The resulting left-right asymmetries for p-shell knockout are compared
in Fig.\ \ref{fig:wally} with results provided by van Orden.
The van Orden calculation produces very strong oscillations in $A_{LT}$
for $p_m \gtrsim 300$ MeV/$c$ that are absent without spinor distortion.
This feature is much weaker in our calculation using the EDAD1 potential,
but when we use the same LF potential a very strong oscillation that is 
similar to that of van Orden is obtained also.
The most important difference between these calculations is the choice of 
optical model, with variations due to vertex function, form factors,
density distribution, and electron distortion being much smaller.
Hence, we also include calculations using the EEI interaction or the 
EDAD1 optical potential in our standard SV model and find an oscillation 
of intermediate amplitude using the EEI potential.
Although we favor the EEI or EDAD1 potentials, 
which give much better descriptions of proton scattering data for these
energies \cite{Kelly91a,Kelly91c,Flanders91},
we must also recognize that $A_{LT}$ for large $p_m$, where the cross
section has fallen by several decades, is probably sensitive to many 
uncertain aspects or approximations in these models.
Other effects which become important at large $p_m$ include: 
ground-state correlations, channel coupling, two-body currents, and
off-shell extrapolation.
Furthermore, gauge and Gordon ambiguities affect both models and are larger
than the differences between them.
Therefore, the relatively good agreement between these two approaches 
suggests that the effective momentum approximation is adequate for 
modest $p_m$.  
Nevertheless, the EMA is not an essential part of the coordinate-space
approach and can be eliminated with sufficient computational investment
should the need arise.

\section{Conclusions}
\label{sec:conclusions}

We have proposed an effective current operator for nucleon electromagnetic
knockout that incorporates spinor distortion by Dirac scalar and vector 
potentials using the effective momentum approximation.
This operator can be used with either relativistic or nonrelativistic
optical potentials and overlap functions, permitting a systematic
investigation of the effects of dynamical enhancement of lower components
and of ambiguities in the off-shell vertex function.
This method can also be used to investigate possible effects of density
dependence in nucleon form factors.
The coordinate-space approach provides a more natural model for investigation
of the possible effects medium modfications of nucleon form factors 
than does the momentum-space approach.

We used this method to study relativistic and density-dependent effects 
upon quasifree $^{16}$O$(\vec{e},e^\prime \vec{p})$ reactions at 
$Q^2 = 0.8$ (GeV/$c$)$^2$,
kinematics relevant to two recent experiments at Jefferson Laboratory.
We find that spinor distortion significantly enhances the left-right
asymmetry $A_{LT}$, but that the magnitude of this enhancement is subject
to a larger Gordon ambiguity than comparable nonrelativistic calculations.
Similarly, the polarization ratio $r_{xz} =  P^\prime_x / P^\prime_z$ for
p-shell knockout is reduced by about 5\% with a $\pm 5\%$ Gordon ambiguity
at the kinematics of experiment E89-033, but the Gordon ambiguity again
increases more rapidly with missing momentum than in nonrelativistic
calculations.
These effects are stronger for the s-shell than for the p-shell.
The most important relativistic effect for $A_{LT}$ comes from the 
bound-state spinor while the dominant relativistic effect for $r_{xz}$ is 
contributed by the ejectile spinor; 
hence, consistency between these quantities should provide a stringent test 
of analyses which seek to extract a medium-modified form factor ratio 
$G_E/G_M$ from quasifree $(\vec{e},e^\prime \vec{p})$ data.

We estimated the possible medium modification of $r_{xz}$ using a recent 
calculation of density-dependent nucleon form factors based upon a 
quark-meson coupling model.
For modest $p_m$ one might expect an additional $8\%$ reduction in
$r_{xz}$ for p-shell proton knockout from $^{16}$O, 
while much larger effects are expected for s-shell knockout.
Therefore, the quasifree $^4$He$(\vec{e},e^\prime \vec{p})$ reaction should
provide a decisive measurement of the nucleon form factor ratio in
the nuclear interior.

\acknowledgements
We thank Prof. J. W. van Orden for providing tables of his calculations
and Prof. D. H. Lu for tables of the QMC form factors.
The support of the U.S. National Science Foundation under grant PHY-9513924 
is gratefully acknowledged.


\begin{thebibliography}{10}

\bibitem{Lu98a}
D.~H. Lu, A.~W. Thomas, K. Tsushima, and A.~G. Williams, Phys. Lett. {\bf {\bf
  B417}},  217  (1998).

\bibitem{Lu98b}
D.~H. Lu, K. Tsushima, A.~W. Thomas, and A.~G. Williams, Nucl. Phys. {\bf {\bf
  A634}},  443  (1998).

\bibitem{Thomas98a}
A.~W. Thomas, D.~H. Lu, K. Tsushima, and A.~G. Williams, {Recent results from
  QMC relevant to TJNAF}, 1998, nucl-th/9807027.

\bibitem{Kelly97}
J.~J. Kelly, Phys. Rev. {\bf C} {\bf {\bf 56}},  2672  (1997).

\bibitem{Arnold81}
R.~G. Arnold, C.~E. Carlson, and F. Gross, Phys. Rev. {\bf C} {\bf {\bf 23}},
  363  (1981).

\bibitem{Kelly99a}
J.~J. Kelly, Phys. Rev. {\bf C} {\bf {\bf 59}},  in press  (1999).

\bibitem{Kelly96}
J.~J. Kelly, Adv.\ Nucl.\ Phys.\ {\bf {\bf 23}},  75  (1996).

\bibitem{Boffi93}
S. Boffi, C. Giusti, and F.~D. Pacati, Phys. Rep. {\bf {\bf 226}},  1  (1993).

\bibitem{Boffi96}
S. Boffi, C. Giusti, F.~D. Pacati, and M. Radici, {\em {Electromagnetic
  Response of Atomic Nuclei}} (Oxford University Press, Oxford, 1996).

\bibitem{Boffi80a}
S. Boffi, C. Giusti, and F.~D. Pacati, Nucl. Phys. {\bf {\bf A336}},  416
  (1980).

\bibitem{Giusti80}
C. Giusti and F.~D. Pacati, Nucl. Phys. {\bf {\bf A336}},  427  (1980).

\bibitem{McVoy}
K.~W. McVoy and L. van Hove, Phys. Rev. {\bf {\bf 125}},  1034  (1962).

\bibitem{Hedayati-Poor95}
M. {Hedayati--Poor}, J.~I. Johansson, and H.~S. Sherif, Phys. Rev. {\bf C} {\bf
  {\bf 51}},  2044  (1995).

\bibitem{Picklesimer85}
A. Picklesimer, J.~W. van Orden, and S.~J. Wallace, Phys. Rev. {\bf C} {\bf
  {\bf 32}},  1312  (1985).

\bibitem{Picklesimer89}
A. Picklesimer and J.~W. van Orden, Phys. Rev. {\bf C} {\bf {\bf 40}},  290
  (1989).

\bibitem{Boffi87}
S. Boffi, C. Giusti, and F.~D. Pacati, Il Nuovo Cimento {\bf {98A}},  291
  (1987).

\bibitem{Jin94a}
Y. Jin and D.~S. Onley, Phys. Rev. {\bf C} {\bf {\bf 50}},  377  (1994).

\bibitem{Udias95}
J.~M. Ud\'{i}as, P. Sarriguren, E.~M. de~Guerra, E. Garrido, and J.~A.
  Caballero, Phys. Rev. {\bf C} {\bf {\bf 51}},  3246  (1995).

\bibitem{Hama90}
S. Hama, B.~C. Clark, E.~D. Cooper, H.~S. Sherif, and R.~L. Mercer, Phys. Rev.
  {\bf C} {\bf {\bf 41}},  2737  (1993).

\bibitem{Horowitz86}
C.~J. Horowitz and B.~D. Serot, Nucl. Phys. {\bf {\bf A368}},  503  (1986).

\bibitem{TIMORA}
C.~J. Horowitz, D.~P. Murdoch, and B.~D. Serot,  in {\em {Computational Nuclear
  Physics I: Nuclear Structure}}, edited by K. Langanke, J.~A. Maruhn, and
  S.~E. Koonin (Springer-Verlag, Berlin, 1991), pp.\ 129--151.

\bibitem{Kelly89b}
J.~J. Kelly, Phys. Rev. {\bf C} {\bf {\bf 39}},  2120  (1989).

\bibitem{Kelly94a}
J.~J. Kelly and S.~J. Wallace, Phys. Rev. {\bf C} {\bf {\bf 49}},  1315
  (1994).

\bibitem{deForest83}
T. de~Forest, Nucl. Phys. {\bf {\bf A392}},  232  (1983).

\bibitem{Chinn92}
C.~R. Chinn and A. Picklesimer, Il Nuovo Cimento {\bf {\bf 105A}},  1149
  (1992).

\bibitem{MMD}
P. Mergell, U.-G. Meissner, and D. Drechsel, Nucl. Phys. {\bf {\bf A596}},  367
   (1996).

\bibitem{TJNAF89-003}
A. Saha {\it et~al.}, {Study of the Quasielastic $(e,e^\prime p)$ Reaction in
  $^{16}$O at High Recoil Momentum, TJNAF Proposal 89-003}, 1989.

\bibitem{TJNAF89-033}
C. Glashausser {\it et~al.}, {Measurement of Recoil Polarization in the
  $^{16}$O$(e,e^\prime p)$ Reaction with 4 GeV Electrons, TJNAF Proposal
  89-033}, 1989.

\bibitem{TJNAF93-049}
J.~F.~J. van~den Brand {\it et~al.}, {Polarization Transfer in the Reaction
  $^4$He$(\vec{e},e^\prime \vec{p})$ in the Quasi-elastic Scattering Region,
  TJNAF Proposal 93-049}, 1993.

\bibitem{Picklesimer87}
A. Picklesimer and J.~W. van Orden, Phys. Rev. {\bf C} {\bf {\bf 35}},  266
  (1987).

\bibitem{LF81}
W.~G. Love and M.~A. Franey, Phys. Rev. {\bf C} {\bf {\bf 24}},  1073  (1981).

\bibitem{Hynes85}
M.~V. Hynes, A. Picklesimer, P.~C. Tandy, and R.~M. Thaler, Phys. Rev. {\bf C}
  {\bf {\bf 31}},  1438  (1985).

\bibitem{Kelly91a}
J.~J. Kelly {\it et~al.}, Phys. Rev. {\bf C} {\bf {\bf 43}},  1272  (1991).

\bibitem{Kelly91c}
J.~J. Kelly {\it et~al.}, Phys. Rev. {\bf C} {\bf {\bf 44}},  2602  (1991).

\bibitem{Flanders91}
B.~S. Flanders {\it et~al.}, Phys. Rev. {\bf C} {\bf {\bf 43}},  2103  (1991).

\end{thebibliography}

\newpage

\begin{figure}[htb] 
\centerline{ \strut\psfig{file=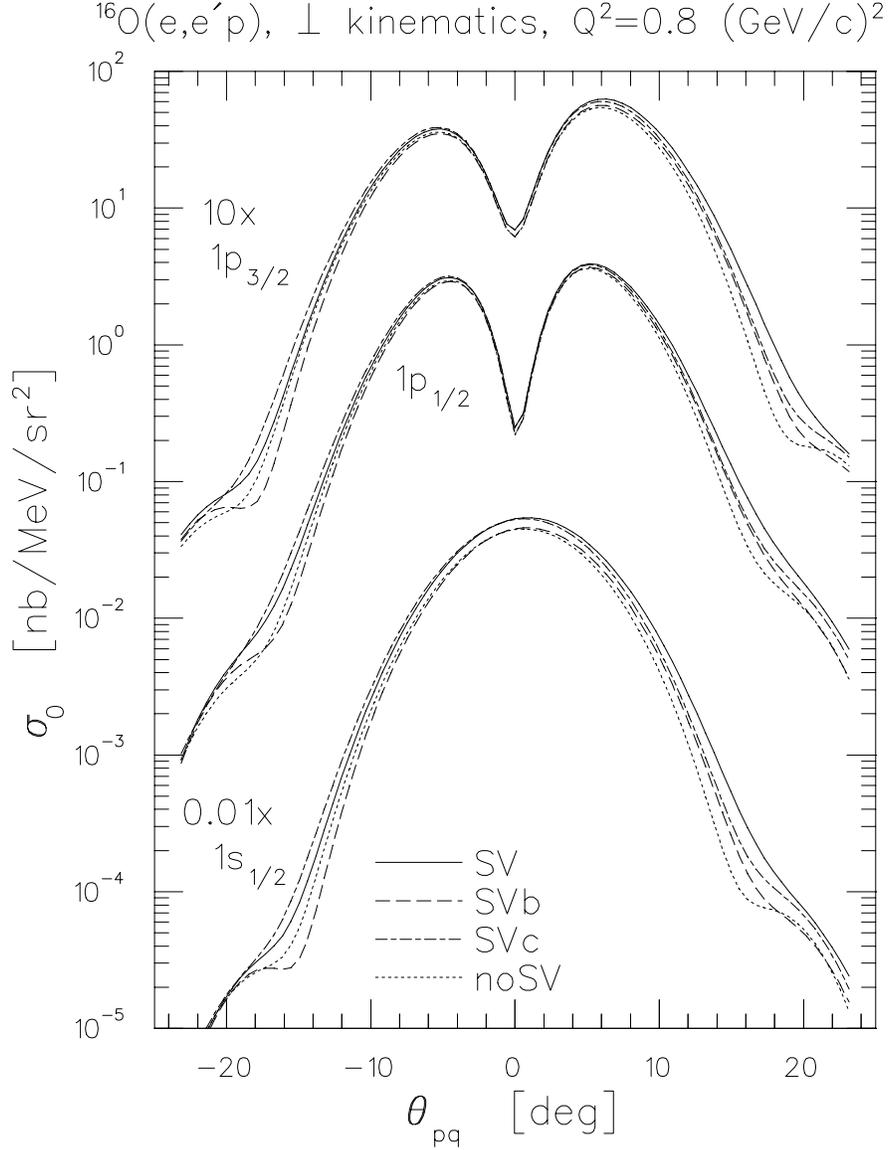,height=6.0in} }
\caption{Calculations of the differential cross section for 
$^{16}$O$(e,e^\prime p)$ in quasiperpendicular kinematics with 
$Q^2 = 0.8$ (GeV/$c$)$^2$ using the $\bar{\Gamma}_1$ current.
Solid curves use the full SV model, dotted curves show noSV,
dashed curves show SVb, and dash-dotted curves SVc calculations.
These calculations are normalized to full subshell occupancy.}
\label{fig:sig_cc1}
\end{figure}

\begin{figure}[htb] 
\centerline{ \strut\psfig{file=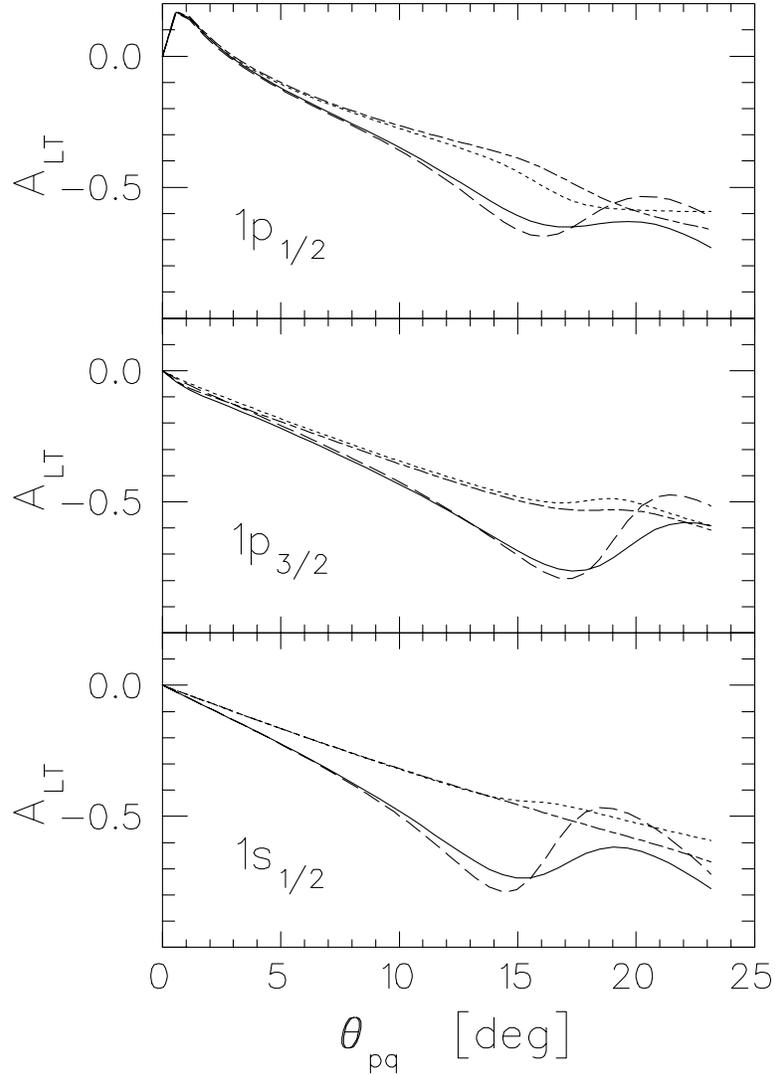,height=6.0in} }
\caption{Calculations of the left-right asymmetry for 
$^{16}$O$(e,e^\prime p)$ in quasiperpendicular kinematics with 
$Q^2 = 0.8$ (GeV/$c$)$^2$ using the $\bar{\Gamma}_1$ current.
Solid curves use the full SV model, dotted curves show noSV,
dashed curves show SVb, and dash-dotted curves SVc calculations.}
\label{fig:alt_cc1}
\end{figure}

\begin{figure}[htb] 
\centerline{ \strut\psfig{file=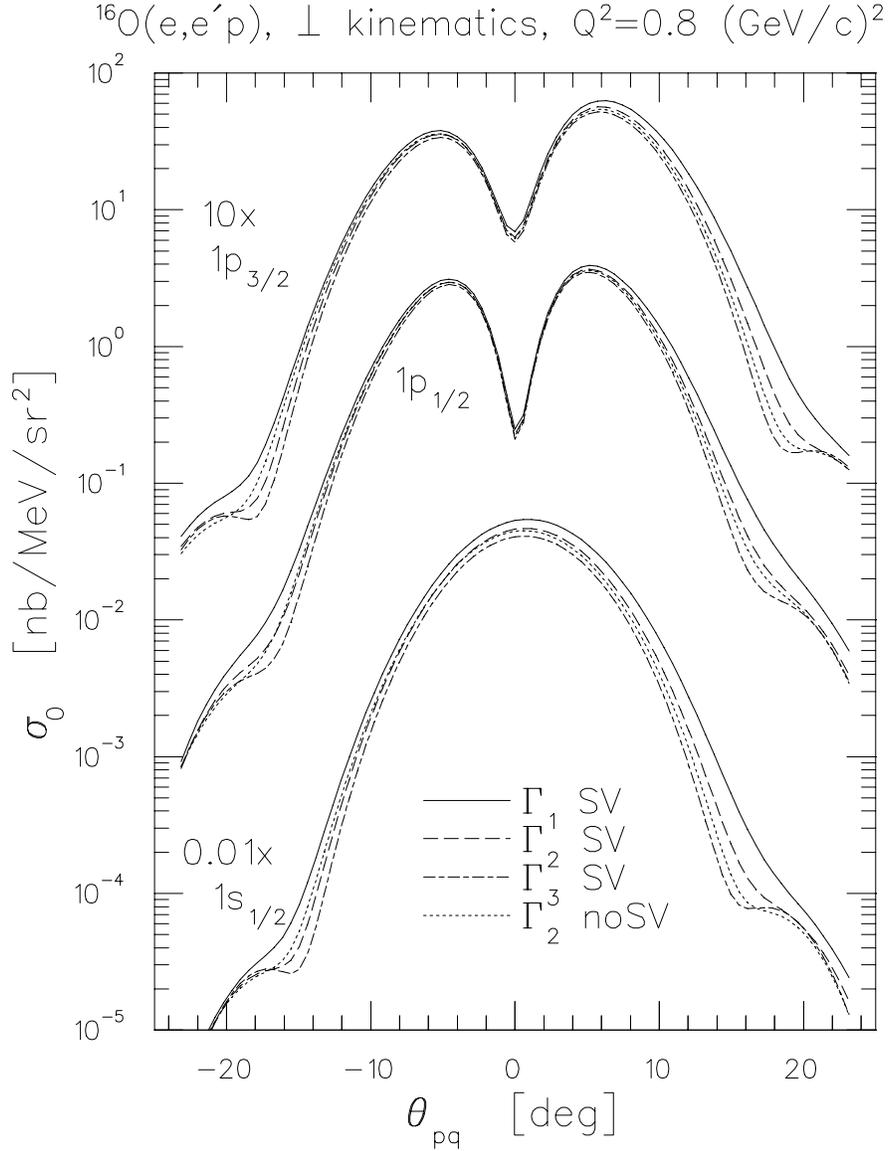,height=6.0in} }
\caption{Calculations of the differential cross section for 
$^{16}$O$(e,e^\prime p)$ in quasiperpendicular kinematics with 
$Q^2 = 0.8$ (GeV/$c$)$^2$.
Solid curves use $\bar{\Gamma}_1$, dashed curves use $\bar{\Gamma}_2$,
and dash-dotted curves use $\bar{\Gamma}_3$ in the full SV model.
Dotted curves use $\bar{\Gamma}_2$ in the noSV model.} 
\label{fig:sig_SV}
\end{figure}

\begin{figure}[htb] 
\centerline{ \strut\psfig{file=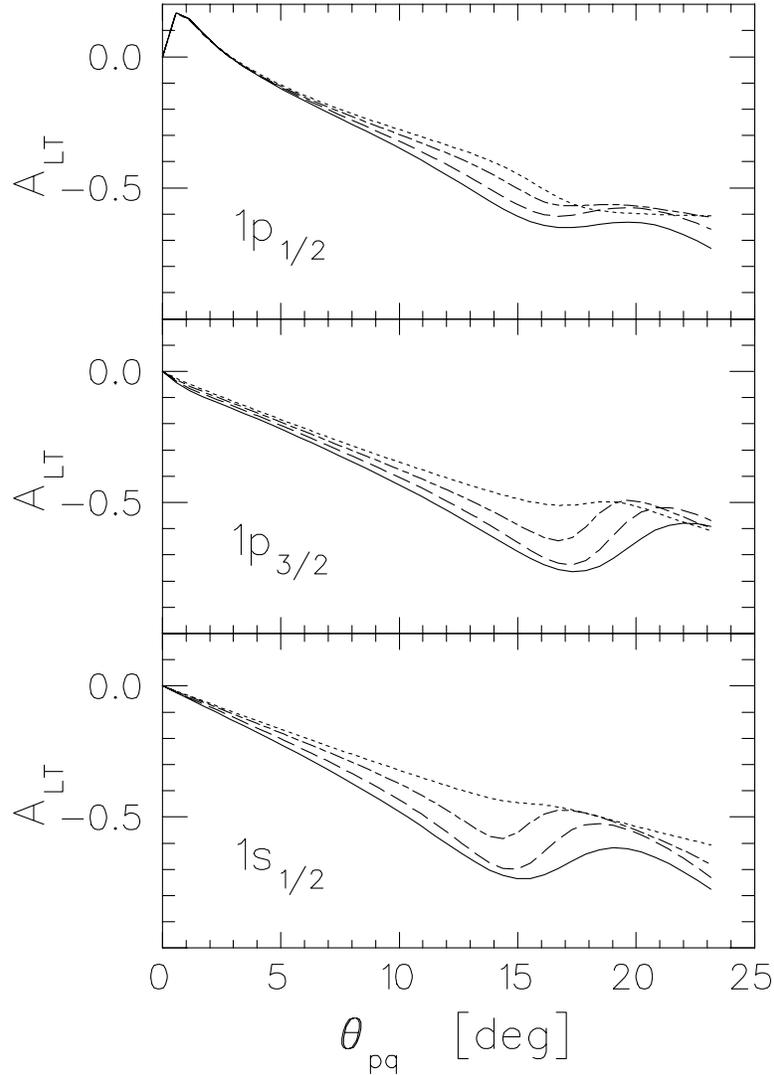,height=6.0in} }
\caption{Calculations of the left-right asymmetry for 
$^{16}$O$(e,e^\prime p)$ in quasiperpendicular kinematics with 
$Q^2 = 0.8$ (GeV/$c$)$^2$. 
Solid curves use $\bar{\Gamma}_1$, dashed curves use $\bar{\Gamma}_2$,
and dash-dotted curves use $\bar{\Gamma}_3$ in the full SV model.
Dotted curves use $\bar{\Gamma}_2$ in the noSV model.}
\label{fig:alt_SV}
\end{figure}

\begin{figure}[htb] 
\centerline{ \strut\psfig{file=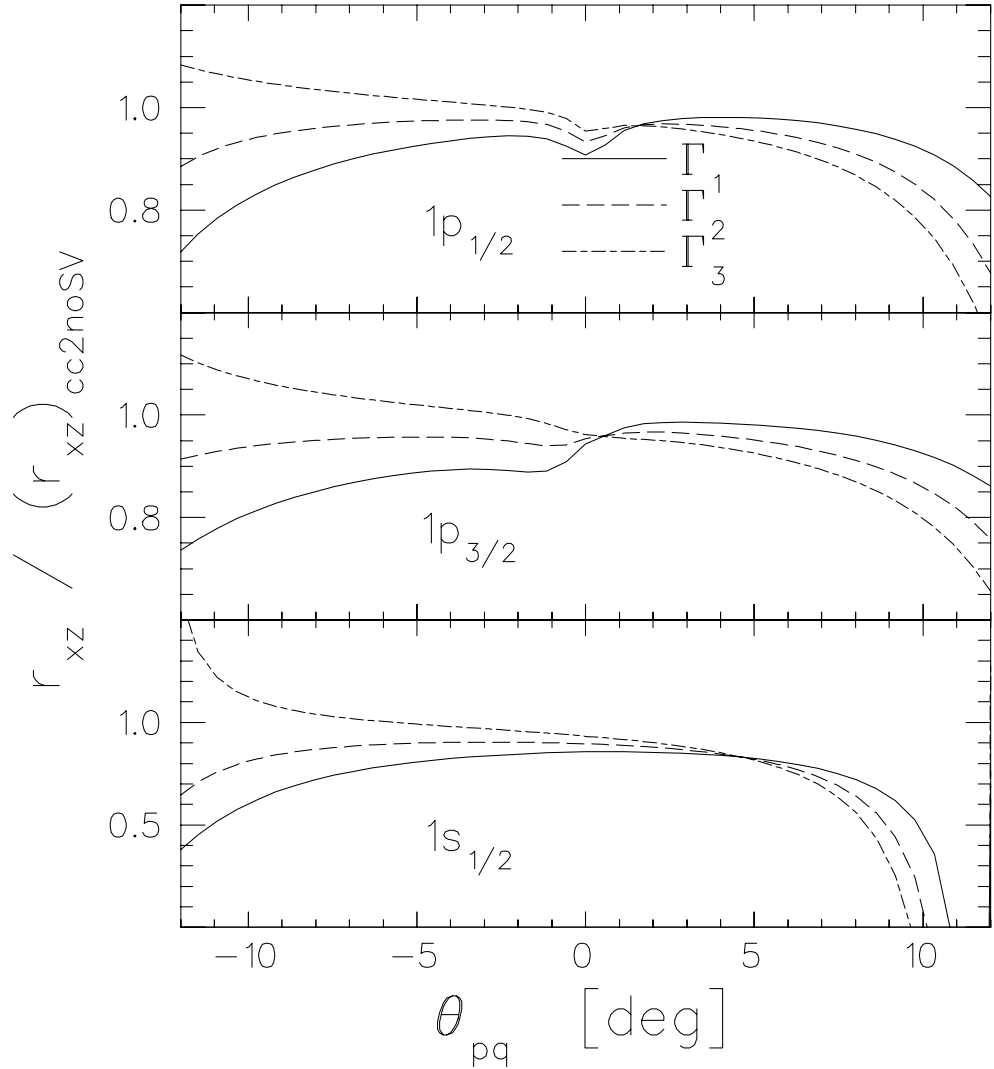,height=6.0in} }
\caption{Calculations of the polarization ratio $r_{xz}$ for 
$^{16}$O$(\vec{e},e^\prime \vec{p})$ in quasiperpendicular kinematics 
with $Q^2 = 0.8$ (GeV/$c$)$^2$ are compared with a baseline calculation
using the $\bar{\Gamma}_2$ current in the noSV model.
Solid curves use $\bar{\Gamma}_1$, dashed curves use $\bar{\Gamma}_2$,
and dash-dotted curves use $\bar{\Gamma}_3$ in the full SV model.}
\label{fig:rxz_SV}
\end{figure}

\begin{figure}[htb] 
\centerline{ \strut\psfig{file=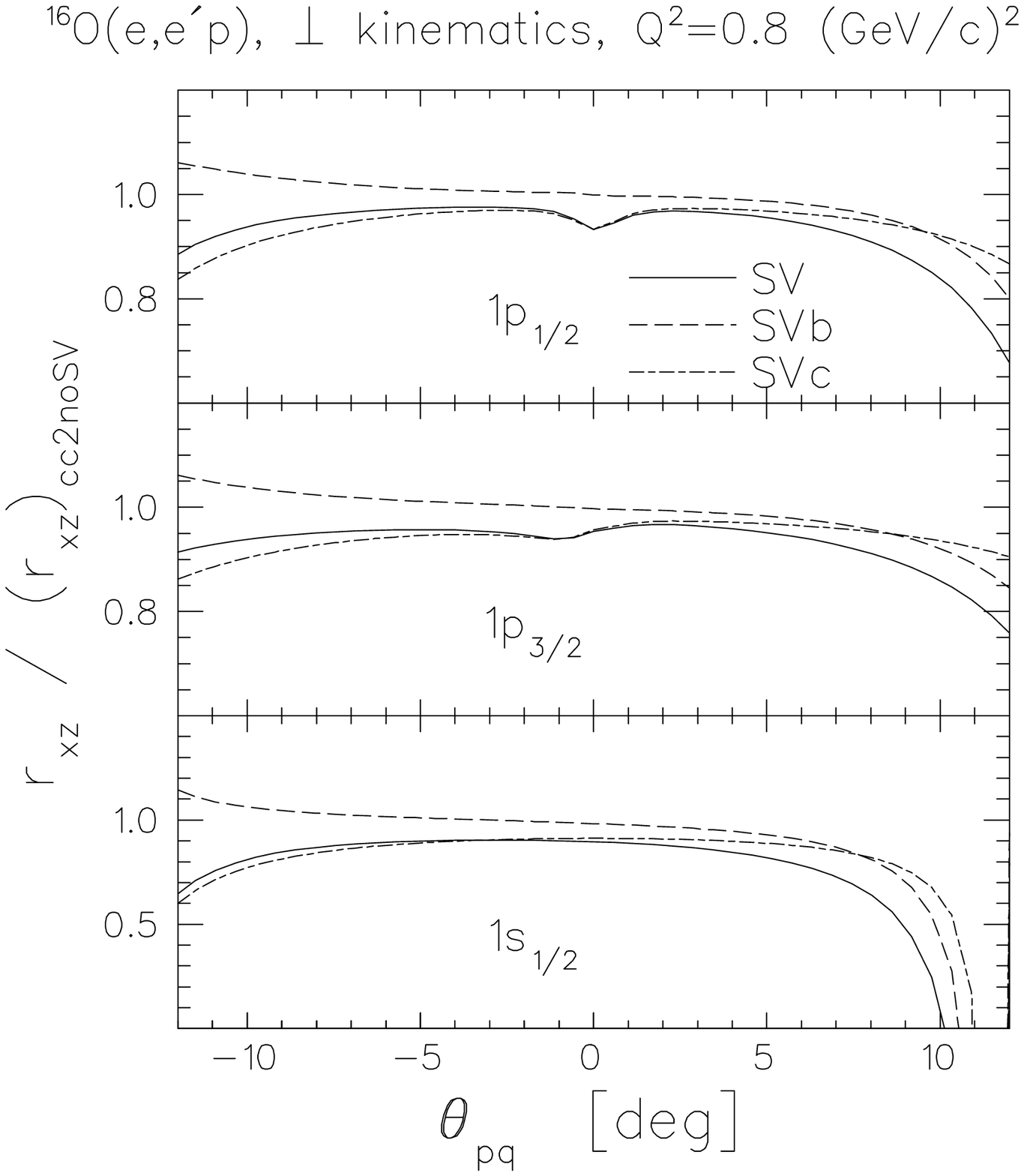,height=6.0in} }
\caption{Calculations of the polarization ratio $r_{xz}$ for 
$^{16}$O$(\vec{e},e^\prime \vec{p})$ in quasiperpendicular kinematics 
with $Q^2 = 0.8$ (GeV/$c$)$^2$ using the $\bar{\Gamma}_2$ current
are compared with a baseline noSV calculation.
Solid curves use the full SV model, dashed curves use SVb,
and dash-dotted curves use SVc.}
\label{fig:rxz_cc2}
\end{figure}

\begin{figure}[htb] 
\centerline{ \strut\psfig{file=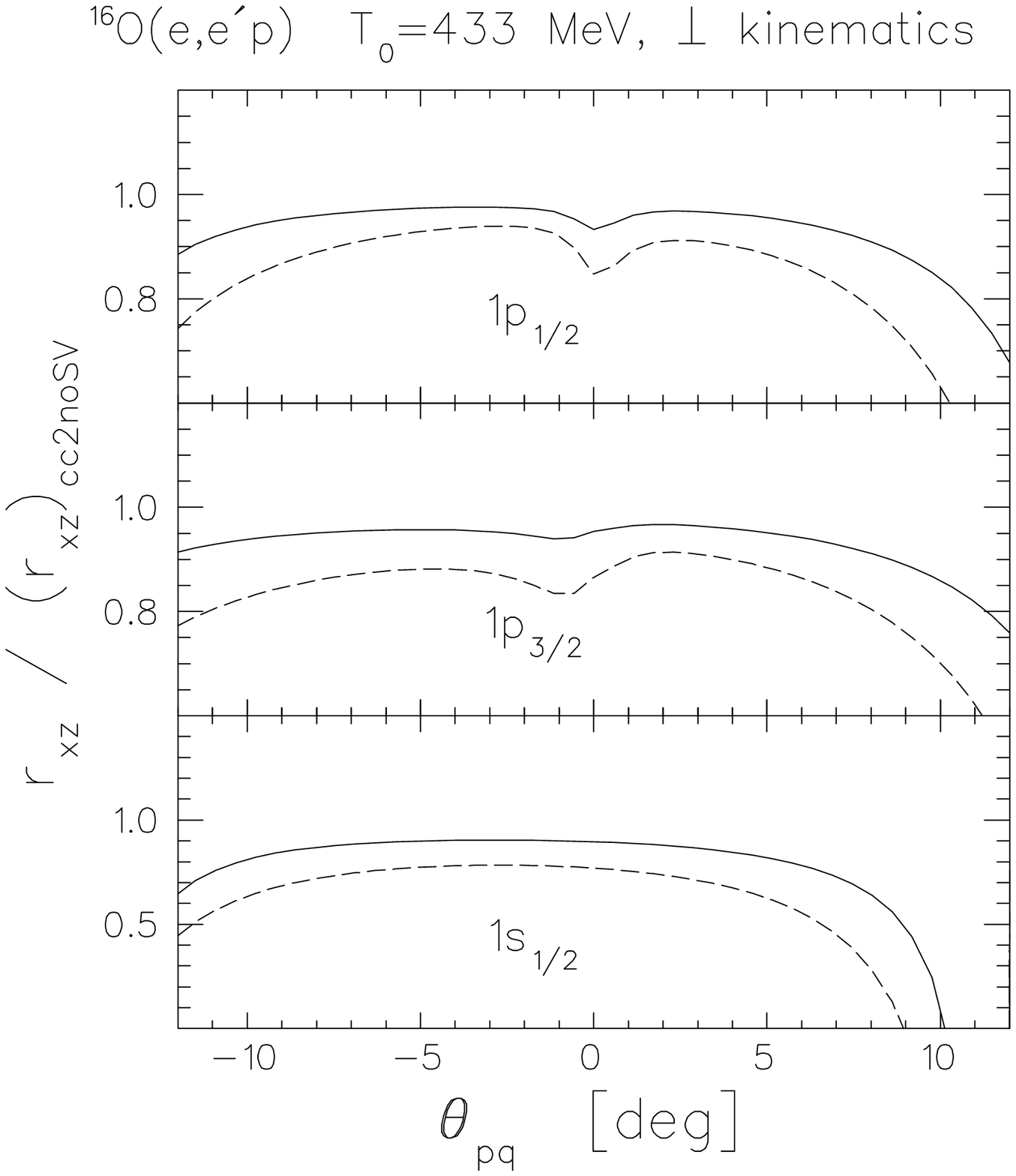,height=6.0in} }
\caption{Calculations of the polarization ratio $r_{xz}$ for 
$^{16}$O$(\vec{e},e^\prime \vec{p})$ in quasiperpendicular kinematics 
with $Q^2 = 0.8$ (GeV/$c$)$^2$ using the $\bar{\Gamma}_2$ current
are compared with a baseline noSV calculation.
Solid curves use the SV model with free form factors while 
dashed curves use the density dependence of QMC form factors.}
\label{fig:rxz_QMC}
\end{figure}

\begin{figure}[htb] 
\centerline{ \strut\psfig{file=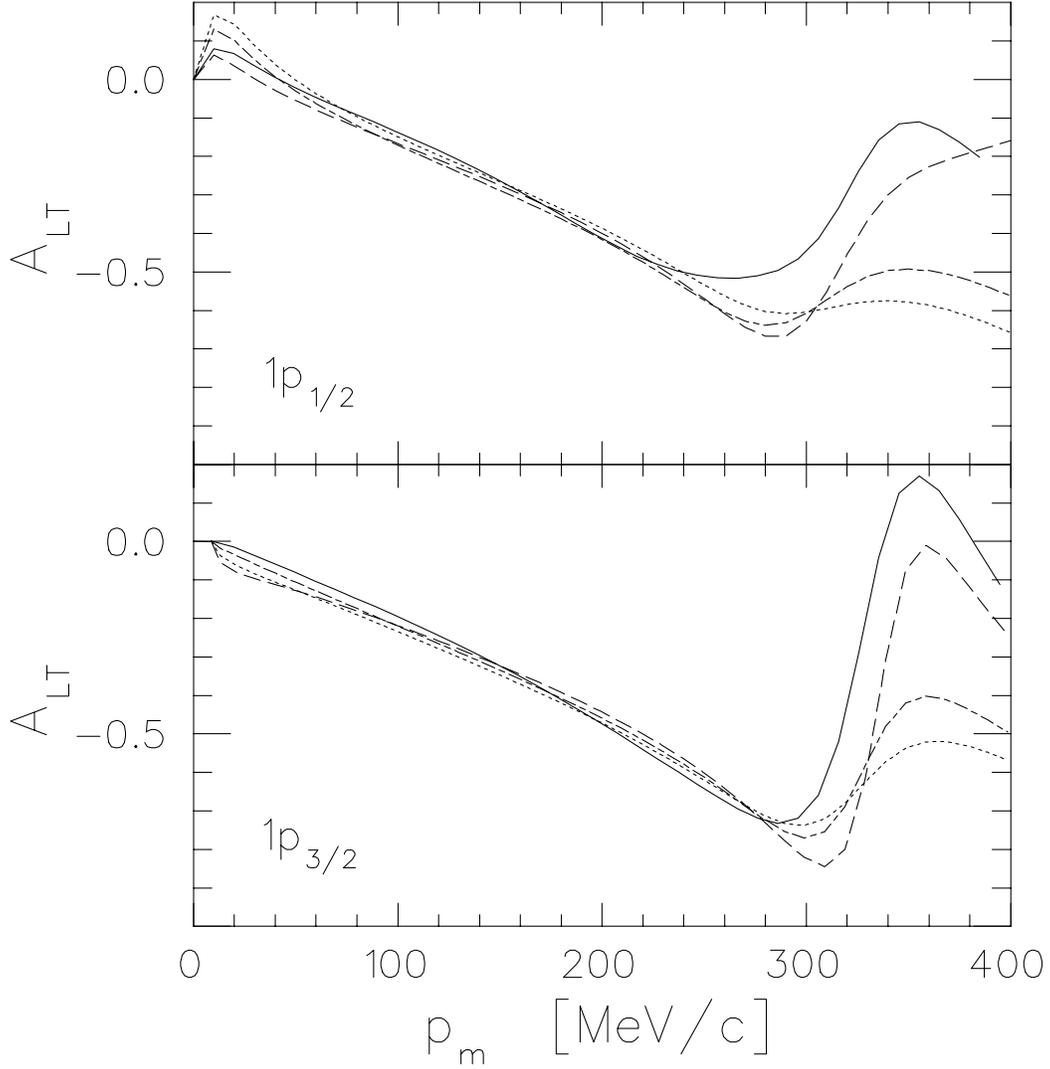,height=6.0in} }
\caption{Calculations of the left-right asymmetry for 
$^{16}$O$(e,e^\prime p)$ in quasiperpendicular kinematics with 
$Q^2 = 0.8$ (GeV/$c$)$^2$.
Solid curves show results provided by van Orden while dashed curves
show our calculations with similar input.
EMA calculations using $\bar{\Gamma}_2$ in the SV model with the
EEI or EDAD1 potentials are shown as dash-dotted or dotted curves,
respectively.}
\label{fig:wally}
\end{figure}

\end{document}